\documentclass[12pt]{article}
\usepackage{geometry}
\usepackage{a4}
\usepackage{graphicx}
\usepackage{epsf}
\usepackage{amsmath}
\usepackage{amssymb}
\usepackage{cite}
\newcommand{\be}{\begin{equation}}
\newcommand{\ee}{\end{equation}}

\newcommand{\Rmnum}[1]{\expandafter\@slowromancap\romannumeral #1@}
\newcommand{\bea}{\begin{eqnarray}}
\newcommand{\eea}{\end{eqnarray}}
\begin{document}
\def\C{{\mathbb{C}}}
\def\R{{\mathbb{R}}}
\def\s{{\mathbb{S}}}
\def\T{{\mathbb{T}}}
\def\Z{{\mathbb{Z}}}
\def\W{{\mathbb{W}}}
\def\Bbb{\mathbb}
\def\BZ{\Bbb Z} \def\BR{\Bbb R}
\def\BW{\Bbb W}
\def\BM{\Bbb M}
\def\BC{\Bbb C} \def\BP{\Bbb P}
\def\CP{\BC\BP}
\begin{titlepage}
\title{R-Charged Black Holes and Holographic Optics} \author{} 
\date{
Prabwal Phukon, Tapobrata Sarkar 
\thanks{\noindent 
E-mail:~ prabwal, tapo@iitk.ac.in} 
\vskip0.4cm 
{\sl Department of Physics, \\ 
Indian Institute of Technology,\\ 
Kanpur 208016, \\ 
India}} 
\maketitle 

\abstract{We analyze momentum dependent vector modes in the context of gauge theories dual to R-charged black holes in $D=4$, $5$ and $7$. 
For a variety of examples, the master variables are constructed, for which the linearized equations for the perturbations decouple. These allow for the
computation of momentum dependent correlation functions. Away from the hydrodynamic limit, numerical analysis using the decoupled equations of
motion is used to obtain the analogues of the Depine-Lakhtakia (DL) index. For specified ranges of frequencies, a negative index of refraction 
is seen to occur in all cases. 
}
\end{titlepage}

\section{Introduction}
\label{intro}
In recent years, there has been increasing evidence that the AdS/CFT correspondence \cite{malda} can been used as a tool to understand realistic physical 
phenomena in material systems. This correspondence, which relates a strongly coupled quantum field theory in flat space-time with a classical theory of gravity in higher 
dimensions, has provided an exciting arena to study conformal field theories that arise in condensed matter systems, via holography. 
Several important and interesting results have thus far emerged, and it has been shown that it is indeed possible to gain useful insight into a variety of  problems 
in condensed matter physics via the gauge-gravity duality \cite{hartnoll},\cite{sachdev1}. 

Important ingredients in this analysis are the retarded correlators of the boundary theory, at finite momentum and frequency. These have been analyzed 
for RN-AdS backgrounds (see e.g \cite{matsuo1},\cite{matsuo2},\cite{leigh}), and provide useful information about the boundary theory. The finite momentum 
analysis can be substantially more difficult than that at zero momentum, due to the appearance of coupled differential equations at the perturbative level, and it 
is useful to decouple the equations by an appropriate choice of ``master variables'' \cite{kodamaishibashi}. These become important in the context of numerical analysis, 
if one needs to go beyond the large wavelength limit in the boundary theory. 

The above discussion becomes relevant in the context of a holographic description of an exotic property of matter that has attracted a lot of attention in recent years, 
namely, the phenomenon of negative refractive index in materials (see, e.g \cite{sar},\cite{valesagorev}). These materials, called meta-materials,
are artificially engineered, and have the property that the permittivity $\epsilon$ and permeability $\mu$ of these can be negative. This results in a negative refractive index,  
in the sense that the negative sign in the square root of the relation $n = -\sqrt{\epsilon\mu}$ had to be chosen \cite{veselago}. Although mathematically
interesting, it took a few decades before the physical applications of the results of \cite{veselago} could be envisaged in \cite{pendry1},\cite{pendry2}. 
Shortly afterwards, the possibility of simultaneous negative values of the $\epsilon$ and $\mu$ was experimentally demonstrated \cite{exp}, and since then,
there has been widespread activity to understand the exotic properties of meta-materials, and these have been used in a variety of applications in physics
and engineering. 

Indeed, in the context of the AdS/CFT correspondence, a holographic description of strongly coupled field theories was shown to
generically admit a negative refractive index \cite{policastro1}. This ``optics/geometry'' duality, studied in the context of 
the five dimensional RN-AdS black hole in \cite{policastro1} has been further explored in \cite{sin1}, where a four dimensional example was
worked out. The methods were then extended to the case of holographic superconductors \cite{hartnoll2}, in the work of \cite{gaozhang}, \cite{policastro2}. 

In this paper, we extend this analysis to gauge theories dual to R-charged black holes in $D=5$, $4$ and $7$, that correspond to rotating $D3$, $M2$ and 
$M5$ branes. In particular, we study the the momentum dependent vector mode perturbations on the black hole side. In several examples, we construct
the master variables, which simplify the computation and gives rise to a set of decoupled equations that result from linearizing the perturbations. 
These are then used to compute a variety of correlators for the boundary theory, which, in particular, leads to the Depine-Lakhtakia (DL) index, that, 
under certain circumstances, determine the possibility of negative refraction in a material medium. While the DL index is simple to write down in the
hydrodynamic limit using results from standard black hole thermodynamics, a numerical routine has to be employed away from this limit. Using our master variables, 
we achieve this, and show that negative refraction is a generic feature in all our examples both in the hydrodynamic limit and away from it, in lines with the 
observations made in \cite{policastro1}. 

This paper is organized as follows. In section 2, we first review the basic setup and fix the notations and conventions to be used in the rest of the paper. 
We then analyze momentum dependent vector mode perturbations corresponding to gauge theories dual to four dimensional R-charged black holes. 
This is done for the single charge case, as well as a two-charge examples. For both cases, we perform numerical computations to 
analyze the DL index away from the hydrodynamic limit. In sections 3 and 4, we perform similar analyses for the five dimensional and the seven dimensional 
examples. Section 5 ends the paper with our conclusions and some discussions. 

\section{R-charged Black Holes in Four Dimensions}

In this section, we will first review some of the basic formulae in order to fix the notations and conventions that we use in the rest of the paper. 
We first quickly recapitulate some of the well known results of R-charged black holes, before reviewing the basic formalism required to understand
their optical properties via holography. Here, and in what follows, our notations closely follow \cite{natsuume}. R-charged black holes form the gravity
duals to rotating branes in various dimensions. For example, in its simplest form, the gravity dual to $D3$-branes is the canonical $AdS_5 \times {\rm S}^5$, 
while a spinning $D3$-brane configuration \cite{gubser} corresponds to adding rotations in planes orthogonal to the brane, which lead to the rotation group
$SO(6)$. Three non-zero chemical potentials can be associated to the three commuting Cartan generators of the group $SO(6)$ of rank $3$, and 
these are the spins of the $D3$-branes, i.e represent the three charges under a global $SO(6)$ R-symmetry of the $N=4$ SCFT residing on the brane. 
These are the three $U(1)$ charges of black holes in $AdS_5$ supergravity that arises in the corresponding Kaluza-Klein reduction of the $D3$-brane on $S^5$
\cite{cvetic}. 

The hydrodynamics of R-charged black holes have been investigated in detail in the last few years, starting from the work of \cite{sonstarinets}. While the
original work of \cite{sonstarinets} focussed on the dual $D3$-brane $N=4$ SYM theories with non-zero chemical potential, this was generalized to other cases
involving $M2$ and $M5$-branes, see e.g \cite{natsuume},\cite{sachin1},\cite{sachin2} and references therein. However, to the best of our
knowledge, momentum dependent analyses of the vector modes for R-charge black holes have not been worked out. In this paper, we will undertake this
task. In what follows, for a variety of examples \footnote{In all the cases considered in this paper, we deal with planar horizons.}, 
we analyze the correlation functions for momentum dependent vector mode perturbations for R-charged 
black holes by constructing a master variable \cite{kodamaishibashi} in each case. In terms of these variables, the perturbation equations decouple, and this
helps us to numerically analyze some properties of the dual gauge theory away from the hydrodynamic limit, as we elaborate upon in sequel. 
Important for our analysis will be the correlators that determine the optical properties of the dual gauge theories. The latter has been investigated in the
pioneering work of \cite{policastro1}. The main result that we will use is that using linear response theory in the AdS/CFT correspondence, it can be shown
\cite{policastro1} that the electric permittivity and the effective magnetic permeability are related to the frequency dependent transverse current correlators
evaluated at the boundary of AdS, and is given by
\begin{eqnarray}
\epsilon(\omega) &=& 1 - \frac{4\pi}{\omega^2}C_{em}^2 G_T^{(0)} \left(\omega\right) \nonumber\\
\mu(\omega) &=& \left[1 + 4\pi C_{em}^2 G_T^{(2)} \left(\omega\right) \right]^{-1} \simeq 1-4\pi C_{em}^{2}G_{T}^{(2)}\left( \omega \right)
\label{epsilonmu}
\end{eqnarray}
where $C_{em}$ is the electromagnetic coupling, and the transverse correlator is expanded as 
\begin{equation}
G_T(\omega, K) =  G_T^{(0)} \left(\omega\right) + K^2G_T^{(2)} \left(\omega\right)
\end{equation}
Negative refraction, which occurs when the phase velocity of the of the light wave is in a direction opposite to its energy flux, 
is then equivalent to the negativity of the DL index \cite{deplak}
\begin{equation}
\eta _{DL}= Re\left(\epsilon \right)| \mu | +Re\left( \mu\right)| \epsilon  |
\label{dl}
\end{equation}
As explained in \cite{policastro1}, $\mu(\omega)$ is an effective magnetic permeability, that is obtained by expanding the transverse dielectric permittivity in a series 
involving the spatial momentum. 

There are several caveats that needs to be kept in mind while analyzing optical properties in the context of the gauge/gravity duality, as explained in \cite{policastro1}.
Firstly, we deal here with relativistic systems, and are somewhat removed from the non relativistic ones usually studied in the laboratory. Secondly, 
strictly speaking, one does not have a dynamical photon in the boundary CFT. Hence we have to imagine a strongly coupled field theory weakly coupled to
such a dynamical electromagnetic field at the boundary, and it is the refractive index of the latter that is being calculated. We will proceed while keeping these caveats in mind. 

\subsection{4-D Single R-charged Black Holes}

We are now ready to analyze the momentum dependent vector modes that correspond to rotating $M2$-branes. We begin with the Lagrangian \cite{natsuume}
\begin{equation}
{\mathcal L} = \sqrt{-g}\left[R-\frac{L^{2}}{8} H^{3/2} F^{2}-\frac{3}{8}\frac{\left( \nabla H\right)^{2} }{H^{2}} +\frac{3}{L^{2}}\left(H^{1/2} +H^{-1/2}\right)  \right] 
\end{equation}
The metric for the planar horizon, and the gauge fields are given by
\begin{eqnarray}
ds_4^{2} &=& \frac{16\left(\pi  {\mathcal T}_0 L \right)^{2} }{9u^{2}}H^{1/2}\left( \frac{-f}{H} dt^{2}+dx^{2}+dz^{2}\right)+\frac{L^{2}}{fu^{2}} H^{1/2}du^{2} 
\nonumber\\
A_{\mu } &=& \frac{4}{3}\pi  {\mathcal T}_0\sqrt{2\kappa \left( 1+\kappa \right) }\frac{u}{H}\left(dt \right) _{\mu }
\end{eqnarray}
As in standard literature, we have defined the coordinate $u = r_+/r$, where $r_+$ is the radius of the outer horizon, $L$ is a scale factor associated with the 
planar part of the metric, and we have defined 
\begin{equation}
H = 1 + \kappa u,~~~ f = \left(1-u \right)\left\lbrace 1 + \left(1 +\kappa  \right)u +  \left(1+\kappa  \right)u^{2} \right\rbrace    
\end{equation}
We also record the expression for the temperature \footnote{The temperature will be denoted by ${\mathcal T}$ in what follows, to keep the notation distinct from
the perturbative analysis},
\begin{equation}
\frac{{\mathcal T}}{{\mathcal T}_{0}} = \frac{1 + \frac{2\kappa }{3}}{\sqrt{1+\kappa }}
\end{equation}
with $\kappa$  denoting the R charge while ${\mathcal T}_{0}$ is the Hawking temperature of neutral black hole.
For our purpose, it is enough to consider perturbations of the metric of the form
\begin{equation}
h_{tx} =	g_{xx}\left( u\right) T\left( u\right) e^{-i\omega t + iKz},  h_{zx} =	g_{xx}\left( u\right) Z\left( u\right) e^{-i\omega t + iKz}, a_{x} = \frac{\mu}{2}A\left( u\right) e^{-i\omega t + iKz}
\end{equation}
Where, $h_{tx}$, $h_{zx}$ and $a_{x}$ are $ \delta g_{tx}$, $ \delta g_{zx}$ and $ \delta A_{x}$ respectively, and all other fluctuations are taken to be zero.
Here, $\mu$ is the chemical potential, and the fluctuation in the gauge field is defined with a factor of $\mu$, to simplify calculations. 
The linearized equations of the perturbations in terms of $T\left(u \right)$, $Z\left(u \right)$ and $A\left(u \right)$ can be written as,
\begin{eqnarray}
&~& T^{'} + \frac{qf}{\varpi H } Z^{'} + \frac{\kappa u^{2}}{2H}A  = 0 \nonumber\\
&~&T^{''} +\frac{uH^{'}-2 H}{u H }T^{'}-\frac{9 q^{2}}{4 f }T-\frac{9 \varpi q}{4 f }Z + \frac{\kappa u^{2}}{2H}A^{'}  = 0 \nonumber\\
&~&Z^{''} + \frac{uf^{'}-2 f}{u f } Z^{'} + \frac{9 \varpi ^{2} H}{4 f^{2}} Z + \frac{9 \varpi q H}{4f^{2} } T = 0\nonumber\\
&~&\left( Hf A^{'} +2\left( \kappa +1\right)T\right)^{'} + \frac{9}{4}\left( \frac{\varpi^{2} H^{2}}{f}- q^{2}H\right)A  = 0
\label{original4d}
\end{eqnarray}
Where, $\varpi = \frac{\omega }{2\pi {\mathcal T}_0}$ and $q = \frac{K  }{2\pi {\mathcal T}_0}$

It can be seen that the first two  of the above set of equations combined with the fourth can be used to reproduce the third one and hence these can be considered to be 
independent. The first and the second equation can be used to eliminate $Z(u)$ and finally, we end up with two coupled differential equations in $A(u)$ and $T(u)$.
We now construct a new set of variables:
\begin{equation}
\Phi_{\pm }\left(u\right)=\frac{H}{u^2}T'\left(u\right) +\frac{\kappa}{2} A\left(u\right) +C_{\pm}\frac{H}{u}A\left(u\right)
\label{main4d}
\end{equation}
Where, we have defined
\begin{equation}
C_{\pm}=\frac{3}{4 }\left(-1\pm \sqrt{1+\frac{q^2\kappa}{\left(1+\kappa\right)}}\right)
\end{equation}
It can be checked that in terms of the new variable $\Phi$, the above set of equations decouple, and reduce to
\begin{equation}
\Phi''_{\pm }\left(u\right) + \left(\frac{f'}{f}+\frac{2}{u}-\frac{H'}{H}\right) \Phi'_{\pm }\left(u\right)+\frac{9}{4 f^2} \left(\varpi^2 H-q^2 f \right)\Phi_{\pm }\left( u\right)+
C_{\pm} \frac{2(1+\kappa)u}{f H} \Phi_{\pm }\left(u\right)=0
\end{equation}
The boundary condition that follows from eqs.(\ref{original4d}) and (\ref{main4d}) can be written as 
\begin{equation}
\lim _{u\rightarrow 0} \left[ u^2\Phi_{\pm }'\left(u\right)-C_{\pm}uHA'\left(u\right)\right] =\frac{9}{4}\left(q^2T^{(0)}+q\varpi Z^{(0)}\right)-C_{\pm}A^{(0)}
\label{bc1}
\end{equation} 
Where we have defined
\begin{equation}
A^{(0)}=\lim _{u\rightarrow 0} A\left(u\right),~~
T^{(0)}=\lim _{u\rightarrow 0} T\left(u\right),~~
Z^{(0)}=\lim _{u\rightarrow 0} Z\left(u\right)
\end{equation} 
Now, writing the master variables in the series expansion of the form 
\begin{equation}
\Phi_{\pm }\left(u\right)=\frac{a_{\pm}}{u}+b_{\pm}{\rm Log}(u)+d_{\pm}+.....
\end{equation}
we see that eq.(\ref{bc1}) reduces to a simpler form
\begin{equation}
\lim _{u\rightarrow 0} \left[ u^2\Phi_{\pm }'\left(u\right)-C_{\pm}uHA'\left(u\right)\right] =-\lim _{u\rightarrow 0} u\Phi_{\pm }\left(u\right)
\end{equation}
We also have to impose the usual incoming boundary condition,
\begin{equation}
\Phi_{\pm}\left(u\right)=f^{-i \frac{\varpi {\mathcal T}_{0}}{2{\mathcal T}}}Y_{\pm}\left(u\right)
\label{ypm4da}
\end{equation}
where, $Y_{\pm}\left(u \right)$ are regular at $u = 1$. Eq.(\ref{main4d}) can then be cast into the form 
\begin{eqnarray}
&~& Y''_{\pm } + \left[ \frac{\left(\frac{ fu^{2}}{H}\right) ^{'}}{\frac{fu^{2}}{H}}-\frac{i\varpi {\mathcal T}_0f^{'}}{{\mathcal T}f}\right]  Y'_{\pm }\nonumber\\
&+& \left[ \frac{9}{4 f^2} \left(\varpi^2 H-q^2 f \right)-\frac{i\varpi {\mathcal T}_0}{2{\mathcal T}}\frac{\left(\frac{f^{'}u^{2}}{H}\right) ^{'}}{\frac{fu^{2}}{H}}
+C_{\pm} \frac{2(1+\kappa)u}{f H}-\frac{\varpi^{2}{\mathcal T}_0^{2} f^{'2}}{4 {\mathcal T}^2 f^2}\right]  Y_{\pm }=0~~~~~~
\label{ypm4d}
\end{eqnarray}
In the hydrodynamic limit, we can solve for $Y_{+}$ as follows. Expanding $Y_{+}\left( u\right) $ in a series of $i\varpi$ and $q^{2}$ as 
 \begin{equation}
Y_{+}\left( u\right) = Y_{+}^{0}\left( u\right)+Y_{+}^{1}\left( u\right)\left( i\varpi\right) + Y_{+}^{2}\left( u\right)\ q^{2}+ ...... 
\end{equation}
we get the following set of equations:
\begin{eqnarray}
&~& Y_{+}^{0''} + \frac{\left(\frac{ fu^{2}}{H}\right)'}{\frac{fu^{2}}{H}}Y_{+}^{0'} =0\nonumber\\
&~& Y_{+}^{1''} + \frac{\left(\frac{ fu^{2}}{H}\right)'}{\frac{fu^{2}}{H}}Y_{+}^{1'} -\frac{{\mathcal T}_0f^{'}}{{\mathcal T}f} Y_{+ }^{0'}
-\frac{ {\mathcal T}_0}{2{\mathcal T}}\frac{\left(\frac{f'u^{2}}{H}\right)'}{\frac{fu^2}{H}} Y_{+ }^0=0
\nonumber\\
&~& Y_{+}^{2''} + \frac{\left(\frac{ fu^{2}}{H}\right)'}{\frac{fu^{2}}{H}}Y_{+}^{2'}+\left[ -\frac{9}{4 f} +\frac{3\kappa u}{4f H}\right]  Y_{+}^0=0
\label{yeq}
\end{eqnarray}
Upon solving the first equation of of the set of eqs.(\ref{yeq}), it can be checked that the regularity condition at $u=1$ fixes $Y_{+}^{0}$  upto a constant.
\begin{equation}
Y_{+}^{0}  = C_p
\end{equation}
Similarly, we obtain  
\begin{equation}
Y_{+}^{1} =C_p \frac{G\left(u \right)}{u}
\end{equation}
\begin{figure}[t!]
\begin{minipage}[b]{0.5\linewidth}
\centering
\includegraphics[width=2.8in,height=2.3in]{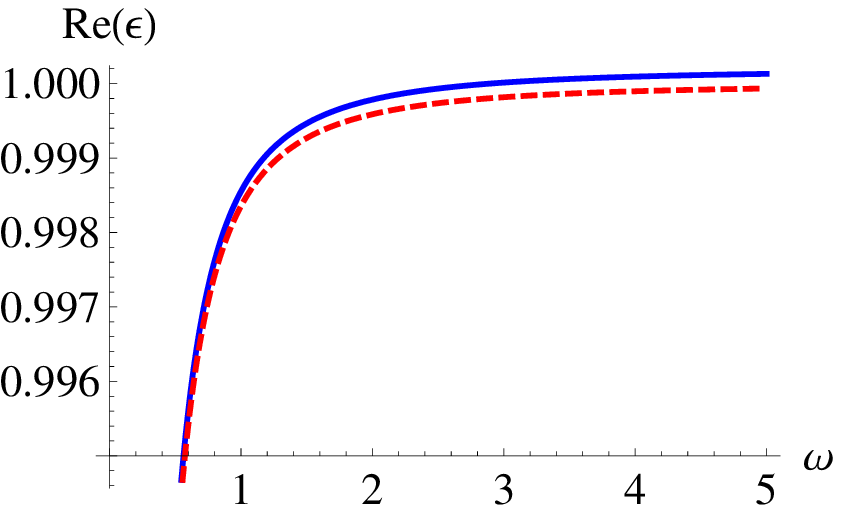}
\caption{Plot of the real part of the permittivity as a function of $\omega$ for 4D Single R-Charge black hole, for $\kappa = 1$. 
The numerical (solid blue line) and analytical (dashed red line) coincide in the hydrodynamic limit.}
\label{compare1}
\end{minipage}
\hspace{0.2cm}
\begin{minipage}[b]{0.5\linewidth}
\centering
\includegraphics[width=2.8in,height=2.3in]{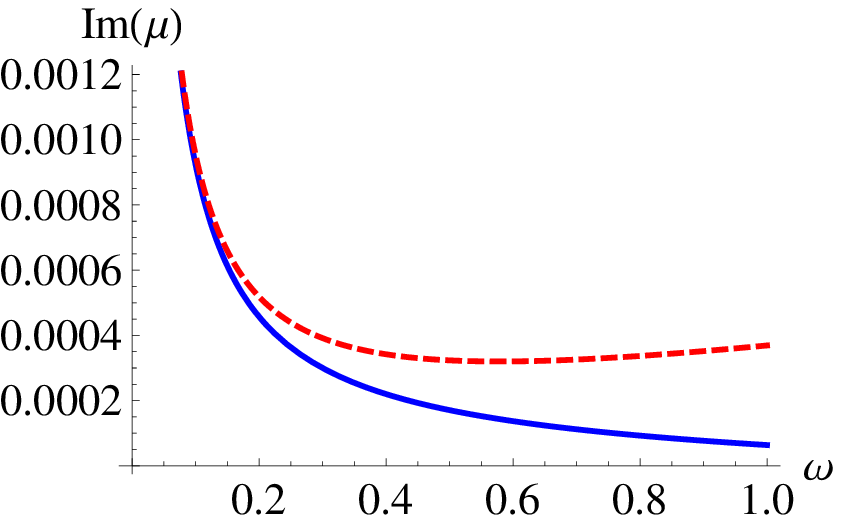}
\caption{Plot of the imaginary part of the effective permeability as a function of $\omega$ for 4D Single R-Charge black hole, for $\kappa = 1$. 
The numerical (solid blue line) and analytical (dashed red line) coincide in the hydrodynamic limit.}
\label{compare2}
\end{minipage}
\end{figure}
Here, in addition to the regularity condition, we have also imposed  the condition, $Y_{+}^{1}(1) =0$. Doing the same to
$Y_{+}^{2} $ we get,
\begin{equation}
Y_{+}^{2} =C_p \frac{H\left(u \right)}{u}
\end{equation}
Finally, the boundary condition (17) gives us $C_p$.
\begin{equation}
C_p =\frac{-\frac{9}{4}\left(q^2T^{(0)}+q\varpi Z^{(0)}\right)+C_{+}A^{(0)}}{i\varpi G(0)+q^2 H(0)}
 \end{equation} 
We record here the closed form expressions for $G(u)$ and $H(u)$. We find that these are given by 
\begin{eqnarray}
&~&G(u) = uY_{+}^{1}(u)~=~\frac{3\left({\mathcal A}+{\mathcal B}\right)}{4\sqrt{1+\kappa}\sqrt{3-\kappa}(3+2\kappa)},~~ H(u)=uY_{+}^{2}(u)=\frac{3(1-u)}{4(1+\kappa)};~~~~~~
\nonumber\\
&~&{\mathcal A}= 2u\sqrt{1+\kappa}(3+\kappa)\left({\rm Tan}^{-1}\left[\frac{3\sqrt{1+\kappa}}{\sqrt{3-\kappa}}\right]-{\rm Tan}^{-1}\left[\frac{(1+2u)
\sqrt{1+\kappa}}{\sqrt{3-\kappa}}\right]\right)~~~~~~
\nonumber\\
&~&{\mathcal B} = \sqrt{3-\kappa}\left(2(u-1)(3+2\kappa)+3u(1+\kappa){\rm Log}\left[\frac{1+u(1+u)(1+\kappa)}{3+2\kappa}\right] \right)~~~~~~
\end{eqnarray}
\begin{figure}[t!]
\begin{minipage}[b]{0.5\linewidth}
\centering
\includegraphics[width=2.8in,height=2.3in]{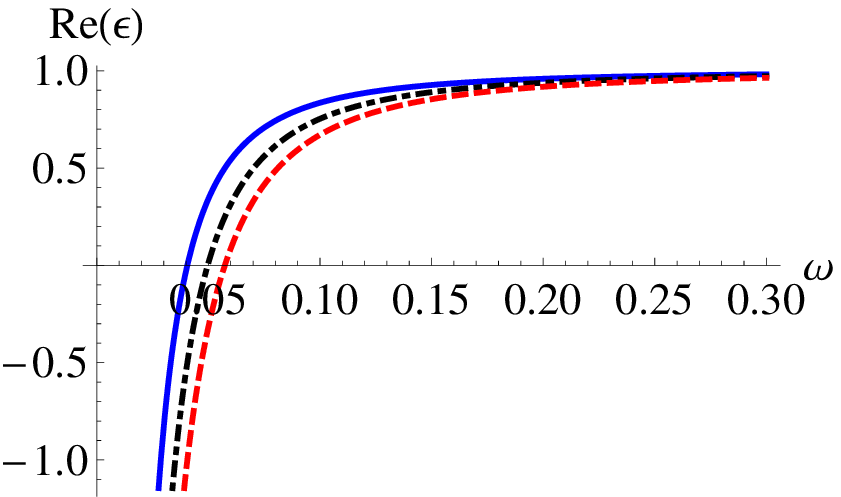}
\caption{Numerical results for the real part of the permittivity as a function of $\omega$, for $\kappa = 1$ (solid blue), $1.5$ (dot dashed black) 
and $2$ (dashed red).}
\label{4dReepsilon}
\end{minipage}
\hspace{0.2cm}
\begin{minipage}[b]{0.5\linewidth}
\centering
\includegraphics[width=2.8in,height=2.3in]{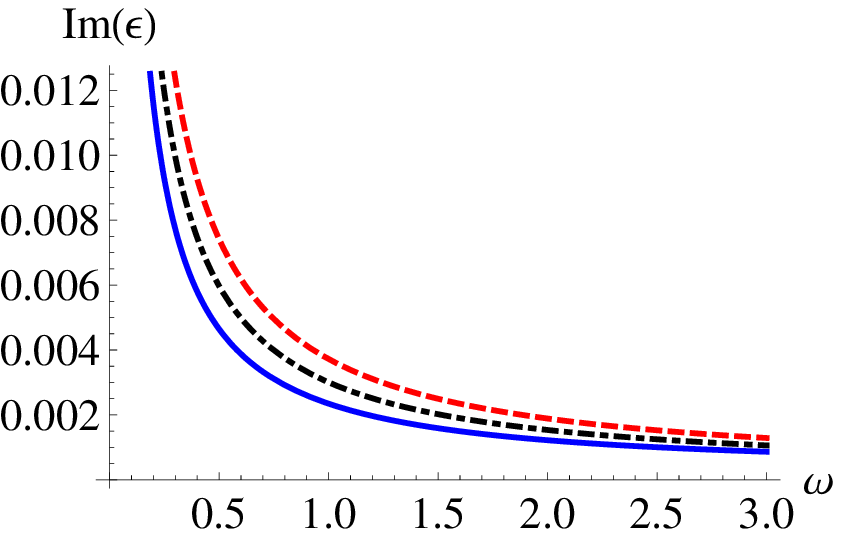}
\caption{Numerical results for the imaginary part of the permittivity as a function of $\omega$, for $\kappa = 1$ (solid blue), $1.5$ (dot dashed black) 
and $2$ (dashed red).}
\label{4dImepsilon}
\end{minipage}
\end{figure}
Now, in the same manner, we solve for $Y_{-}$. First, we expand, $Y_{-}\left( u\right) $ in a series of $i\varpi$ and $q^{2}$.
 \begin{equation}
Y_{-}\left( u\right) = Y_{-}^{0}\left( u\right)+Y_{-}^{1}\left( u\right)\left( i\varpi\right) + Y_{-}^{2}\left( u\right)\ q^{2}+ ...... 
\end{equation}
Subjecting the solutions to appropriate boundary conditions as before, we get 
\begin{equation}
Y_{-}^{0}  = \frac{3+2u\kappa}{2u\kappa}C_n, ~~Y_{-}^{1} = \frac{3+2u\kappa}{2u\kappa}C_n P\left(u \right),~~
Y_{-}^{2} =\frac{3+2u\kappa}{2u\kappa}C_n Q\left(u \right)
\end{equation}
Where, we have defined
\begin{equation}
C_n =\frac{2\kappa}{3}\left[ \frac{C_{-}A^{(0)}-\frac{9}{4}\left(q^2T^{(0)}+q\varpi Z^{(0)}\right)}{1+i\varpi P(0)+q^2 Q(0)}\right] 
\end{equation} 
The closed form expressions for $P(u)$ and $Q(u)$ are presented here for the sake of completeness : 
\begin{eqnarray}
P(u)&=& \frac{\left({\mathcal C}+{\mathcal D}\right)}{4(3-\kappa)^{3/2}\sqrt{1+\kappa}(3+2\kappa)};\nonumber\\
Q(u) &=& \frac{6(u-1)\sqrt{3-\kappa}\kappa^2+27\sqrt{1+\kappa}(3+2u\kappa){\mathcal R}}{2(3-\kappa)^{3/2}(1+\kappa)(3+2\kappa u)}; \nonumber\\
{\mathcal C} &=& 6(\kappa^2+12\kappa+9)\sqrt{1+\kappa}\left({\rm Tan}^{-1}\left[\frac{(1+2u)\sqrt{1+\kappa}}{\sqrt{3-\kappa}}\right]-
{\rm Tan}^{-1}\left[\frac{3\sqrt{1+\kappa}}{\sqrt{3-\kappa}}\right]\right)
\nonumber\\
{\mathcal D} &=&\sqrt{3-\kappa}\left( \frac{24(1-u)\kappa^3}{3+2u\kappa}+9(\kappa^2-2\kappa-3){\rm Log}\left[\frac{3+2\kappa}{1+u(1+u)(1+\kappa)}\right]\right) 
\nonumber\\
{\mathcal R} &=& \left({\rm Tan}^{-1}\left[\frac{3\sqrt{1+\kappa}}{\sqrt{3-\kappa}}\right]
-{\rm Tan}^{-1}\left[\frac{(1+2u)\sqrt{1+\kappa}}{\sqrt{3-\kappa}}\right]\right)
\end{eqnarray}
Now that we have solved for the $Y_{\pm}$, we can write down the master variables $\Phi_{\pm}$ from eq.(\ref{ypm4da}). An appropriate combination of 
eq.(\ref{main4d}) now yields a solution for $A(u)$. This can again be fed back into eq.(\ref{main4d}) to obtain $T'(u)$, and finally, using the first equation
of eq.(\ref{original4d}), we can calculate $Z'(u)$. It is now easy to calculate the boundary action from ($S_{ct}$ is a counter-term) 
\begin{eqnarray}
S_{B} ~=&~&\lim _{u\rightarrow 0} \int dt \vec{dx}\left[\frac{N_{c}^{3/2}{\mathcal T}_0}{36\sqrt{2}} fHA^{'}A +\frac{8\pi^{2}N_{c}^{3/2}{\mathcal T}_0^3}{81\sqrt{2}}\left(\frac{f}{u^2}Z Z{'}
-\frac{H}{u^2}TT{'} \right)\right.\nonumber\\
&~& \left. +\frac{\pi {\mathcal T}_0^2 N_{c}^{3/2}\sqrt{2\kappa\left( 1+\kappa\right) }}{27\sqrt{2}}TA + S_{ct}\right] 
\end{eqnarray}
and taking its derivatives yield the required Green's functions. 
\begin{figure}[t!]
\begin{minipage}[b]{0.5\linewidth}
\centering
\includegraphics[width=2.8in,height=2.3in]{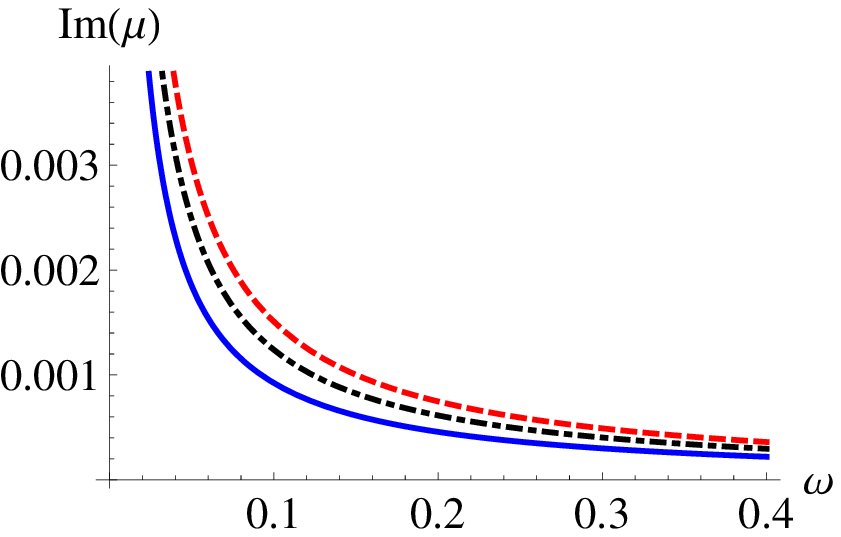}
\caption{Numerical results for the imaginary part of the effective permeability as a function of $\omega$, for $\kappa = 1$ (solid blue), $1.5$ (dot dashed black) 
and $2$ (dashed red).}
\label{4dImmu}
\end{minipage}
\hspace{0.2cm}
\begin{minipage}[b]{0.5\linewidth}
\centering
\includegraphics[width=2.8in,height=2.3in]{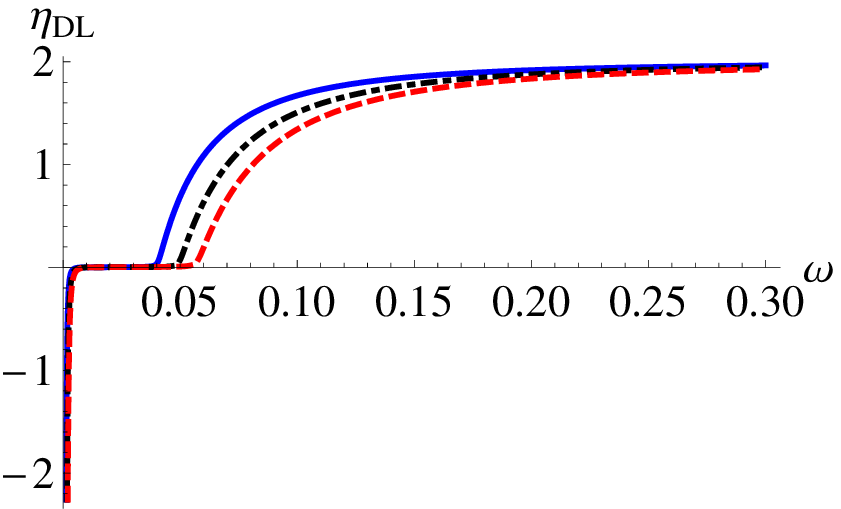}
\caption{Numerical results for the DL index, as a function of $\omega$, for $\kappa = 1$ (solid blue), $1.5$ (dot dashed black) 
and $2$ (dashed red).}
\label{4dDL}
\end{minipage}
\end{figure}
We list the set of retarded correlators in the hydrodynamic limit :
\begin{eqnarray}
G_{xx}&=&\frac{i\omega N_{c}^{3/2}{\mathcal T}_{0} \kappa }{54\sqrt{2}\left(i\omega-\frac{K^2}{4\pi {\mathcal T}_{0}\sqrt{1+\kappa}} \right)}-\frac{i\omega N_{c}^{3/2}
 (3+2\kappa)^2 }{216\sqrt{2}\pi\sqrt{1+\kappa}}
\nonumber\\
G_{xtxt}&=&\frac{4\pi N_{c}^{3/2}{\mathcal T}_{0}^{2}\sqrt{1+\kappa}K^2}{27\sqrt{2}\left(i\omega-\frac{K^2}{4\pi {\mathcal T}_{0}\sqrt{1+\kappa}} \right)},~~
G_{xzxz}=\frac{4\pi N_{c}^{3/2}{\mathcal T}_0^{2}\sqrt{1+\kappa}\omega^2}{27\sqrt{2}\left(i\omega-\frac{K^2}{4\pi {\mathcal T}_0\sqrt{1+\kappa}} \right)}
\nonumber\\
G_{xtxz}&=&-\frac{4\pi N_{c}^{3/2}{\mathcal T}_{0}^{2}\sqrt{1+\kappa}\omega K}{27\sqrt{2}\left(i\omega-\frac{K^2}{4\pi {\mathcal T}_{0}\sqrt{1+\kappa}} \right)},~~
G_{xtx}=\frac{2i\pi N_{c}^{3/2}{\mathcal T}_0^{2}\sqrt{\kappa(1+\kappa)}\omega}{27\left(i\omega-\frac{K^2}{4\pi {\mathcal T}_0\sqrt{1+\kappa}} \right)}\nonumber\\
G_{xzx} &=&\frac{iK}{4\pi {\mathcal T}_0 \sqrt{1+\kappa}}G_{xtx}
\end{eqnarray}
The expression for DL index is obtained from eq.(3), along with the expressions for $\epsilon$ and $\mu$ given in eq.(1). For $|n^2|\ll \frac{1}{\omega D}$, where $
n=\frac{K}{\omega}$ is the refractive index and D is the coffecient in the diffusion pole, we find that upto leading order,
 \begin{equation}
{\epsilon}=1-\frac{\sqrt{2}\pi C_{em}^2 N_{c}^{3/2}{\mathcal T}_0\kappa}{27\omega^{2}}+\frac{i\sqrt{2} C_{em}^2 N_{c}^{3/2}\left(3+2\kappa\right)^2 }{108\omega\sqrt{1+\kappa}}+.....
\end{equation}
\begin{equation}
{\mu}=1+\frac{i C_{em}^2 N_{c}^{3/2}\kappa}{54\sqrt{2}\omega \sqrt{1+\kappa}}+....
\end{equation}
At small enough $\omega$, real part of $\epsilon$ goes to negative values ensuring the DL index to obtain a negative value. 
The cut-off $\omega$ for negative refractive index will vary with the value of the charge parameter $\kappa$. To go beyond the hydrodynamic limit we need
to solve the equations numerically. Let us elaborate on this briefly. The numerical solution away from the hydrodynamic limit is constructed as follows.
We start with the expansion 
\begin{equation}
Y_{\pm} = {\bar Y_{\pm}^0} + q^2 {\bar Y_{\pm}^2}
\end{equation}
Next, we assume the asymptotic expansion at the boundary 
\begin{eqnarray}
{\bar Y_{\pm}^0} &=& \frac{{\hat a_{\pm}}}{u} + {\hat b_{\pm}}~{\rm log}\frac{u}{\delta} + {\hat c_{\pm}} + {\hat d_{\pm}}u \cdots \nonumber\\
{\bar Y_{\pm}^2} &=& \frac{{\hat p_{\pm}}}{u} + {\hat q_{\pm}}~{\rm log}\frac{u}{\delta} + {\hat r_{\pm}} + {\hat s_{\pm}}u \cdots 
\end{eqnarray}
where the hatted quantities are constants. Now, we use these expansions in the equations for ${\bar Y_{\pm}}^0$ and ${\bar Y_{\pm}}^2$ in eq.(\ref{ypm4d})
and obtain a series expansion in terms of the variable $u$. We then need to set the coefficients of $1/u$ and the constant term to zero. Since ${\hat a_{\pm}}$ and ${\hat p_{\pm}}$ are 
known, these can be used to express ${\hat b_{\pm}}$ and ${\hat c_{\pm}}$ in terms of ${\hat d_{\pm}}$ (and ${\hat q_{\pm}}$, ${\hat r_{\pm}}$ in terms of ${\hat e_{\pm}}$).
Once this is done, the constants ${\hat d_{\pm}}$ and ${\hat s_{\pm}}$ are obtained by demanding that the solutions go over to known hydrodynamic case when in the large 
wavelength limit. To illustrate the point, we plot, in figs.(\ref{compare1}) and (\ref{compare2}), results for the numerical  and analytical computation, for 
the real part of the permittivity and the imaginary part of the effective permeability, as a function of $\omega$. These are seen to coincide in the hydrodynamic limit,
and deviate away from it. We emphasize here that knowing the master variables allow us to perform similar analyses for all the retarded correlators. We will, 
in particular, focus on the optical properties of the boundary theory. The results of our numerical analysis for the real and imaginary parts of $\epsilon$ are presented in 
figs.(\ref{4dReepsilon}) and (\ref{4dImepsilon}). In fig.(\ref{4dImmu}), we show the imaginary part of $\mu$, and numerical results for the DL index is shown in fig.(\ref{4dDL}). 
\begin{figure}[t!]
\begin{minipage}[b]{0.5\linewidth}
\centering
\includegraphics[width=2.8in,height=2.3in]{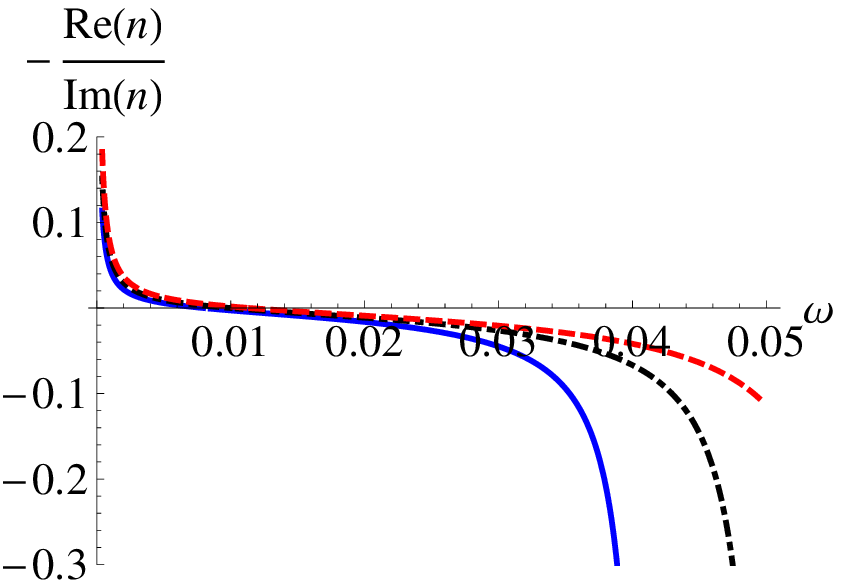}
\caption{$-\frac{Re[n]}{Im[n]}$ as a function of $\omega$, for
$\kappa=1$ (solid blue), 1.5 (dot dashed black) and 2 (dashed red).}
\label{fig7}
\end{minipage}
\hspace{0.2cm}
\begin{minipage}[b]{0.5\linewidth}
\centering
\includegraphics[width=2.8in,height=2.3in]{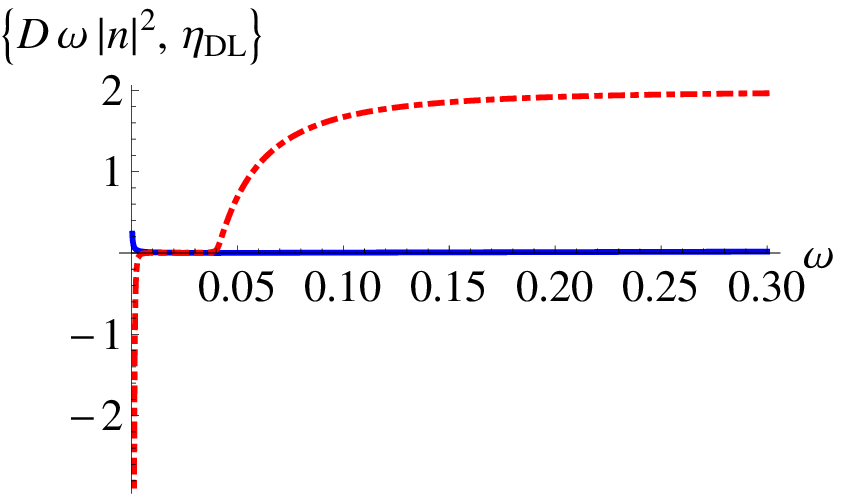}
\caption{$\omega D |n^2|$ (solid blue line) and $\eta_{DL}$ (dashed red line) as
functions of $\omega$, for $\kappa=1$.}
\label{fig8}
\end{minipage}
\end{figure}
In fig.(\ref{fig7}), we show the plot of the ratio $-\frac{Re[n]}{Im[n]}$ to demonstrate the
dissipation effects. At the region of negative refractive index, this ratio is quite
small, as expected for isotropic metamaterials. As we lower $\omega$, dissipation
effects get reduced. On the other hand, at higher frequencies, dissipation becomes
more prominent. The plot is quite similar in nature to that observed in \cite{policastro1}.     

It is to be noted that the validity of the constraint $|n^2|\ll \frac{1}{\omega D}$
has been explicitly verified for all the above results. We have always worked in a
frequency range that falls between the minimal frequency(lower cut-off) and the
critical frequency(higher cut off). To illustrate this claim, we plot $\omega D
|n^2| $  and $\eta_{DL}$ together for $\kappa=1$ in fig.(\ref{fig8}). One can see that,
within the plotted frequency range, $\omega D |n^2| \ll 1 $  holds true. 

\subsection{A Two-charge Example in 4-D} 

We now consider a two-charge example in four dimensions. Here, for simplicity, we will restrict to the case of two equal charges. We begin with the Lagrangian
\begin{equation}
{\mathcal L} = \sqrt{-g}\left[R-\frac{L^2}{8} H^{1/2} \left(F_{1}^{2}+ F_{2}^{2}\right)-\frac{1}{8}\frac{\left( \nabla H\right)^{2} }{H^{2}}+... \right] 
\end{equation}
The metric is given as
\begin{equation}
ds_4^{2} = \frac{16\left(\pi  {\mathcal T}_0 L \right)^{2} }{9u^2}H^{1/2}\left( \frac{-f}{H} dt^{2}+dx^{2}+dz^{2}\right)+\frac{L^{2}}{fu^{2}} H^{1/2}du^{2} 
\end{equation}
and the expressions for the gauge fields become
\begin{eqnarray}
A^{1}_{\mu } &=& \frac{4\pi}{3} {\mathcal T}_0\sqrt{2\kappa} \left( 1+\kappa \right)\frac{u}{\sqrt{H}}\left(dt \right) _{\mu }\nonumber\\
A^{2}_{\mu } &=& \frac{4\pi}{3}  {\mathcal T}_0\sqrt{2\kappa} \left( 1+\kappa \right)\frac{u}{\sqrt{H}}\left(dt \right) _{\mu }
\end{eqnarray}
Where, we define
\begin{equation}
H = H_{1}H_{2}=\left( 1 + \kappa u\right) ^2,~~~f = H-\left( 1+\kappa\right)^2 u^3,~~~\frac{{\mathcal T}}{{\mathcal T}_0} =\frac{3+\kappa}{3}
\end{equation}
$\kappa$  denotes R charge while ${\mathcal T}_0$ denotes the Hawking temperature of neutral black hole.
We consider perturbations of the form 
\begin{eqnarray}
h_{tx} &=&g_{xx}\left( u\right) T\left( u\right) e^{-i\omega t + iKz}, ~~ h_{zx} =	g_{xx}\left( u\right) Z\left( u\right) e^{-i\omega t + iKz}\nonumber\\
a^{1}_{x} &=& \frac{\mu^1 }{2}A^1\left( u\right) e^{-i\omega t + iKz}, ~~a^{2}_{x} = \frac{\mu^2 }{2}A^2\left( u\right) e^{-i\omega t + iKz}
\end{eqnarray}
The linearized perturbation equations are given in terms of $T\left(u \right)$, $Z\left(u \right)$, $A^1\left(u \right)$and $A^2\left(u \right)$,
\begin{eqnarray}
&~&T^{'} + \frac{qf}{\varpi H } Z^{'} + \frac{\kappa\left(1+\kappa\right)u^2}{2H}\left(A^1+A^2\right)  = 0
\nonumber\\
&~&T^{''} +\frac{uH^{'}-2H}{u H }T^{'}-\frac{9q^{2}}{4f }T-\frac{9\varpi q}{ 4f }Z + \frac{\kappa\left(1+\kappa\right)u^2}{2H}\left({A^1}^{'}+{A^2}^{'}\right)  = 0
\nonumber\\
&~&Z^{''} + \frac{uf^{'}-2f}{u f } Z^{'} + \frac{9\varpi ^{2} H}{4f^{2}} Z + \frac{9\varpi q H}{4f^{2}} T = 0
\nonumber\\
&~&{A^1}^{''} + \frac{f^{'}}{f } {A^1}{'} + \frac{2\left(1+\kappa\right)}{f} {T}{'}+\left(\frac{9\varpi ^{2} H}{4f^{2}}-\frac{9q^2}{4f}\right)A^1 = 0
\nonumber\\
&~&{A^2}^{''} + \frac{f^{'}}{f } {A^2}{'} + \frac{2\left(1+\kappa\right)}{f} {T}{'}+\left(\frac{9\varpi ^{2} H}{4f^{2}}-\frac{9q^2}{4f}\right)A^2 = 0
\end{eqnarray}
Where,$\varpi = \frac{\omega }{2\pi {\mathcal T}_0}$ and $q = \frac{K  }{2\pi {\mathcal T}_0}$
The master variables in this case are defined by 
\begin{equation}
\Phi_{\pm }\left(u\right)=\frac{H}{u^2}T'\left(u\right) +\frac{\kappa(1+\kappa)}{2} \left(A^1\left(u\right)+A^2\left(u\right)\right) +C_{\pm}\frac{\sqrt{H}}{u}\left(A^1\left(u\right)+A^2\left(u\right)\right) 
\end{equation}
with the definition of $C_{\pm}$ as 
\begin{equation}
C_{\pm}=\frac{3(1+\kappa)}{8}\left(-1  \pm \sqrt{1+\frac{2q^2\kappa}{\left(1+\kappa\right)^2}}\right)
\end{equation}
In terms of the master variables, the perturbation equations decouple, and we obtain 
\begin{eqnarray}
&~&\Phi''_{\pm }\left(u\right) + \left(\frac{f'}{f}+\frac{2}{u}-\frac{H'}{H}\right) \Phi'_{\pm }\left(u\right)+\frac{9}{4 f^2} \left(\varpi^2 H-q^2 f \right)\Phi_{\pm }\left( u\right)\nonumber\\
&~& ~~~~~ +4C_{\pm} \frac{(1+\kappa)u}{f \sqrt{H}} \Phi_{\pm }\left(u\right)=0 \nonumber\\
&~&\left(A^1-A^2\right)^{''} + \frac{f'}{f}\left(A^1-A^2\right)+\frac{9}{4 f^2} \left(\varpi^2 H-q^2 f\right)\left(A^1-A^2\right)=0
\end{eqnarray}
Now, using the boundary conditions 
\begin{eqnarray}
&~&\lim _{u\rightarrow 0} \left[ u^2\Phi_{\pm }'\left(u\right)-C_{\pm}u\sqrt{H}\left( {A^1}^{'}+{A^2}^{'}\right)\right] =\frac{9}{4}\left(q^2T^{(0)}+q\varpi Z^{(0)}\right) -C_{\pm}\left({A^1}^{0}+{A^2}^{0}\right) 
\nonumber\\
&~&{A^i}^{(0)}=\lim _{u\rightarrow 0} {A^i}\left(u\right),~~T^{(0)}=\lim _{u\rightarrow 0} T\left(u\right),~~Z^{(0)}=\lim _{u\rightarrow 0} Z\left(u\right)
\end{eqnarray} 
we proceed in the same way as the single charge example, and after some algebra, we can write the boundary action, and calculate the transverse correlators.
We find that these are given by
\begin{eqnarray}
G_{xx}^{11}&=&\frac{i\omega N_{c}^{3/2}{\mathcal T}_0 \kappa }{54\sqrt{2}\left(i\omega-\frac{K^2}{4\pi {\mathcal T}_0(1+\kappa)} \right)}-\frac{i\omega N_{c}^{3/2} (9+12\kappa+5\kappa^2) }{216\sqrt{2}\pi(1+\kappa)}=G_{xx}^{22}\nonumber\\
G_{xx}^{12}&=&\frac{i\omega N_{c}^{3/2}{\mathcal T}_0 \kappa }{54\sqrt{2}\left(i\omega-\frac{K^2}{4\pi {\mathcal T}_0(1+\kappa)} \right)}+\frac{i\omega N_{c}^{3/2} \kappa(3+2\kappa) }{108\sqrt{2}\pi(1+\kappa)}=G_{xx}^{21}\nonumber\\
G_{xtxt}&=&\frac{4\pi N_{c}^{3/2}{\mathcal T}_0^{2}(1+\kappa)K^2}{27\sqrt{2}\left(i\omega-\frac{K^2}{4\pi {\mathcal T}_0(1+\kappa)} \right)},~~
G_{xzxz}=\frac{4\pi N_{c}^{3/2}{\mathcal T}_0^{2}(1+\kappa)\omega^2}{27\sqrt{2}\left(i\omega-\frac{K^2}{4\pi {\mathcal T}_0(1+\kappa)} \right)}
\nonumber\\
G_{xtxz}&=&-\frac{4\pi N_{c}^{3/2}{\mathcal T}_0^{2}(1+\kappa)\omega K}{27\sqrt{2}\left(i\omega-\frac{K^2}{4\pi {\mathcal T}_0(1+\kappa)} \right)},~~
G_{xtx}^1=\frac{2i\pi N_{c}^{3/2}{\mathcal T}_0^{2}\sqrt{\kappa}(1+\kappa)\omega}{27\left(i\omega-\frac{K^2}{4\pi {\mathcal T}_0(1+\kappa)}\right)}=G_{xtx}^2\nonumber\\
G_{xzx}^1 &=&-\frac{N_{c}^{3/2}{\mathcal T}_0 \sqrt{\kappa}\omega K}{54\left(i\omega-\frac{K^2}{4\pi {\mathcal T}_0(1+\kappa)}\right)}=G_{xzx}^2
\end{eqnarray}
The components of permittivity and permeability are computed from $G_{xx}^1$and $G_{xx}^2$ using the following relationship:
\begin{equation}
G_{xx}^{ij}=G_{xx0}^{ij}+K^2 G_{xx2}^{ij},~~
{\epsilon}^{ij}=1-\frac{4\pi C_{em}^2}{\omega^{2}} G_{xx0}^{ij},~~
{\mu}^{ij}=1-4\pi C_{em}^2 G_{xx2}^{ij}
\end{equation} 
And, they are found to be 
\begin{eqnarray}
{\epsilon}^{11}&=&1-\frac{\sqrt{2}\pi C_{em}^2 N_{c}^{3/2}{\mathcal T}_0\kappa}{27\omega^{2}}+\frac{i\sqrt{2} C_{em}^2 N_{c}^{3/2}\left(9+12\kappa+5\kappa^2\right) }{108\left(1+\kappa\right)\omega}+.....={\epsilon}^{22}\nonumber\\
{\epsilon}^{12}&=&1-\frac{\sqrt{2}\pi C_{em}^2 N_{c}^{3/2}{\mathcal T}_0\kappa}{27\omega^{2}}-\frac{i\sqrt{2} C_{em}^2 N_{c}^{3/2}\kappa\left(3+2\kappa\right) }{54\left(1+\kappa\right)\omega}+.....={\epsilon}^{21}\nonumber\\
{\mu}^{11}&=&1+\frac{i C_{em}^2 N_{c}^{3/2}\kappa}{54\sqrt{2}\left(1+\kappa\right)\omega }+.....={\mu}^{22}={\mu}^{12}={\mu}^{21}
\end{eqnarray} 
The components of the DL index are given by
\begin{equation}
\eta_{DL}^{ij}=Re[\epsilon^{ij}]|\mu^{ij}|+Re[\mu^{ij}]|\epsilon^{ij}|
\end{equation} 
Again, similar to the previous subsection, we can analyze these in the hydrodynamic limit and away from it. The results of our computations indicate that they
have similar features as that of the single charge example. The only difference is that the imaginary part of $\epsilon^{12}$ seems to be negative, in contrast to
all other cases considered in this paper. 
Note that the case under consideration involves R-charged black holes in 4-D with two equal charges. So, there will be two charge densities $\rho_{1}$, $\rho_{2}$ and two
conjugate chemical potentials $\mu_{1}$, $\mu_{2}$. Kubo's formula, then lead us to a total of 4 components of retarded current-current correlator. These components can
be thought of as the elements of a $2\times 2$ symmetric matrix (similar cases have been studied in \cite{sachin1}). In a similar spirit, one can construct 
symmetric $2\times 2$ matrices for the components of permittivity, permeability and DL index. This is indicative of the fact that the system can, in principle, show 
different responses to a dynamical electromagnetic field weakly coupled at the boundary. A similar comment applies to a two charge example in five dimensions, 
that we consider in sequel. It would be interesting to study this further.

\section{R charged Black Holes in Five Dimensions}

In this section, we will study the optical properties of R-charged black holes in $D=5$, which are dual to rotating $D3$-branes. We will first study the 
single charge example. Although this was worked out in \cite{sonstarinets}, we construct here the master variables which decouple the perturbation 
equations, and allow us to go beyond the hydrodynamic limit, in studying the optical properties. We show this calculation in some details, to 
contrast the method with the well known work of \cite{sonstarinets}. We will then work out a two-charge example. 

\subsection{5-D Single R-charged Black holes}

For the single R-charged black hole in five dimensions, we begin with the Lagrangian
\begin{equation}
{\mathcal L} = \sqrt{-g}\left[R-\frac{L^{2}}{8} H^{4/3} F^{2}-\frac{1}{3}\frac{\left( \nabla H\right)^{2} }{H^{2}} +\frac{4}{L^{2}}\left(H^{2/3} +2 H^{-1/3}\right)  \right] 
\end{equation}
The metric and gauge fields are given by 
\begin{eqnarray}
ds_5^{2} &=& \frac{\left(\pi  {\mathcal T}_0 L \right)^{2} }{u}H^{1/3}\left( \frac{-f}{H} dt^{2}+dx^{2}+dy^{2}+dz^{2}\right)+\frac{L^{2}}{4fu^{2}} H^{1/3}du^{2} 
\nonumber\\
A_{\mu } &=& \pi  {\mathcal T}_0\sqrt{2\kappa \left( 1+\kappa \right) }\frac{u}{H}\left(dt \right) _{\mu }
\end{eqnarray}
Where we define
\begin{equation}
H = 1 + \kappa u,~~~ f = \left(1-u \right)\left\lbrace 1 + \left(1 +\kappa  \right)u  \right\rbrace,~~~\frac{T}{{\mathcal T}_0} = \frac{1 + \frac{\kappa }{2}}{\sqrt{1+\kappa }}
\end{equation}
with $\kappa$ denoting the R-charge while ${\mathcal T}_0$ denotes the Hawking temperature of neutral black hole. Considering perturbations of the form,
\begin{equation}
h_{tx} =	g_{xx}\left( u\right) T\left( u\right) e^{-i\omega t + iKz},  h_{zx} =	g_{xx}\left( u\right) Z\left( u\right) e^{-i\omega t + iKz}, a_{x} = \frac{\mu }{2}A\left( u\right) e^{-i\omega t + iKz}
\end{equation}
with all other fluctuations set to zero, the linearized equations in terms of $T\left(u \right)$, $Z\left(u \right)$ and $A\left(u \right)$ can be written as
\begin{eqnarray}
&~&T^{'} + \frac{qf}{\varpi H } Z^{'} + \frac{\kappa u}{2H}A  = 0
\nonumber\\
&~&T^{''} +\frac{uH^{'}- H}{u H }T^{'}-\frac{q^{2}}{fu }T-\frac{\varpi q}{ fu }Z + \frac{\kappa u}{2H}A^{'}  = 0
\nonumber\\
&~&Z^{''} + \frac{uf^{'}-f}{u f } Z^{'} + \frac{\varpi ^{2} H}{f^{2}u} Z + \frac{\varpi q H}{f^{2}u } T = 0
\nonumber\\
&~&\left( Hf A^{'} +2\left( \kappa +1\right)T\right)^{'} + \left( \frac{\varpi^{2} H^{2}}{uf}- \frac{q^{2}H}{u}\right)A  = 0
\end{eqnarray}
Here, $\varpi = \frac{\omega }{2\pi {\mathcal T}_0}$ and $q = \frac{K  }{2\pi {\mathcal T}_0}$. Now, we construct the generalized variables 
\begin{equation}
\Phi_{\pm }\left(u\right)=\frac{H}{u}T'\left(u\right) +\frac{\kappa}{2} A\left(u\right) +C_{\pm}\frac{H}{u}A\left(u\right)
\end{equation}
where we have defined 
\begin{equation}
C_{\pm}=\frac{1}{2 }\left(-1\pm \sqrt{1+\frac{q^2\kappa}{\left(1+\kappa\right)}}\right)
\end{equation}
and obtain the equations
\begin{equation}
\Phi''_{\pm }\left(u\right) + \left(\frac{f'}{f}+\frac{2}{u}-\frac{H'}{H}\right) \Phi'_{\pm }\left(u\right)+\frac{1}{u f^2} \left(\varpi^2 H-q^2 f \right)\Phi_{\pm }\left( u\right)+C_{\pm} \frac{2(1+\kappa)}{f H} \Phi_{\pm }\left(u\right)=0
\label{main5d}
\end{equation}
The boundary conditions can be obtained as in the previous case. Also, imposing 
\begin{equation}
\Phi_{\pm}\left(u\right)=f^{-i \frac{\varpi {\mathcal T}_0}{2T}}Y_{\pm}\left(u\right)
\end{equation}
where, $Y_{\pm}\left(u \right)$ are regular at $u = 1$, their equations reduce to 
\begin{eqnarray}
&~& Y''_{\pm } + \left[ \frac{\left(\frac{ fu^{2}}{H}\right) ^{'}}{\frac{fu^{2}}{H}}-\frac{i\varpi {\mathcal T}_0f^{'}}{Tf}\right]  Y'_{\pm } \nonumber\\
&~& +\left[ \frac{1}{u f^2} \left(\varpi^2 H-q^2 f \right)-\frac{i\varpi {\mathcal T}_0}{2T}\frac{\left(\frac{f^{'}u^{2}}{H}\right) ^{'}}
{\frac{fu^{2}}{H}}+C_{\pm} \frac{2(1+\kappa)}{f H}-\frac{\varpi^{2}{\mathcal T}_0^{2} f^{'2}}{4 T^2 f^2}\right]  Y_{\pm }=0 ~~~~~~
\end{eqnarray}
As before, we expand $Y_{+}\left( u\right) $ in a series of $i\varpi$ and $q^{2}$,
 \begin{equation}
Y_{+}\left( u\right) = Y_{+}^{0}\left( u\right)+Y_{+}^{1}\left( u\right)\left( i\varpi\right) + Y_{+}^{2}\left( u\right)\ q^{2}+ ...... 
\end{equation}
So that we get a series of equations:
\begin{eqnarray}
&~& Y_{+}^{0''} + \frac{\left(\frac{ fu^{2}}{H}\right)'}{\frac{fu^{2}}{H}}Y_{+}^{0'} =0
\nonumber\\
&~& Y_{+}^{1''} + \frac{\left(\frac{ fu^{2}}{H}\right)'}{\frac{fu^{2}}{H}}Y_{+}^{1'} -\frac{{\mathcal T}_0f^{'}}{Tf} Y_{+ }^{0'}-\frac{ {\mathcal T}_0}{2T}\frac{\left(\frac{f'u^{2}}{H}\right)'}{\frac{fu^2}{H}} Y_{+ }^0=0
\nonumber\\
&~& Y_{+}^{2''} + \frac{\left(\frac{ fu^{2}}{H}\right)'}{\frac{fu^{2}}{H}}Y_{+}^{2'}+\left[ -\frac{1}{u f} +\frac{\kappa u}{2f H}\right]  Y_{+}^0=0
\end{eqnarray}
Proceeding as in the previous cases, we obtain 
\begin{equation}
Y_{+}^{0}  = C_p,~~
Y_{+}^{1} =C_p \frac{G\left(u \right)}{u},~~~Y_{+}^{2} =C_p \frac{H\left(u \right)}{u},~~~
C_p =\frac{-q^2T^{(0)}-q\varpi Z^{(0)}+C_{+}A^{(0)}}{i\varpi G(0)+q^2 H(0)}
\end{equation} 
\begin{figure}[t!]
\begin{minipage}[b]{0.5\linewidth}
\centering
\includegraphics[width=2.8in,height=2.3in]{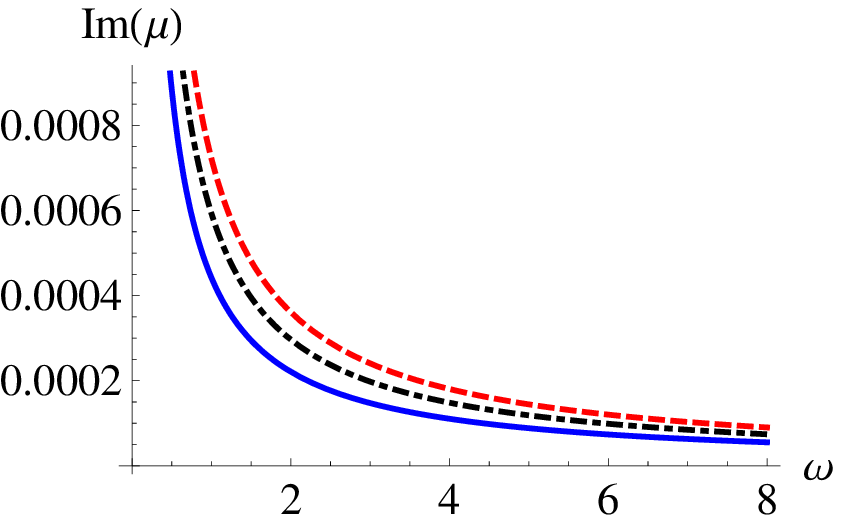}
\caption{Numerical results (5D single charge case) for the imaginary part of the effective permeability as a function of $\omega$, for $\kappa = 1$ (solid blue), $1.5$ (dot dashed black) 
and $2$ (dashed red).}
\label{5dImmu}
\end{minipage}
\hspace{0.2cm}
\begin{minipage}[b]{0.5\linewidth}
\centering
\includegraphics[width=2.8in,height=2.3in]{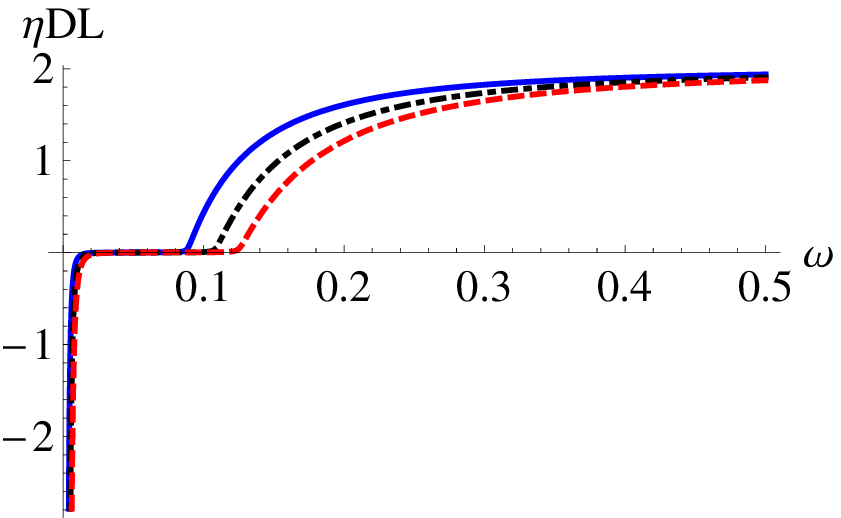}
\caption{Numerical results (5D single charge case) for the DL index, as a function of $\omega$, for $\kappa = 1$ (solid blue), $1.5$ (dot dashed black) 
and $2$ (dashed red). This is negative for small $\omega$.}
\label{5dDL}
\end{minipage}
\end{figure}
Here, we have used the expressions
\begin{eqnarray}
G(u)&=&uY_{+}^{1}(u)=\frac{{\mathcal T}_0\left((u-1)(2+\kappa)+2u(1+\kappa){\rm Log}\left[\frac{1+u+u\kappa}{1+\kappa}\right]\right)}{2T(1+\kappa)}
\nonumber\\
H(u)&=&uY_{+}^{2}(u)=\frac{(1-u)}{2(1+\kappa)}
\end{eqnarray}
The solution for $Y_{-}$ can be obtained similarly. Expanding $Y_{-}\left( u\right) $ in a series of $i\varpi$ and $q^{2}$,
and subjecting the solutions to appropriate boundary conditions, it can be checked that 
\begin{equation}
Y_{-}^{0}  = \frac{2+u\kappa}{2u}C_n,~~
Y_{-}^{1} = \frac{2+u\kappa}{2u}C_n P\left(u \right),~~
Y_{-}^{2} =\frac{2+u\kappa}{2u}C_n Q\left(u \right)
\end{equation}
where $C_n$ is obtained as 
\begin{equation}
C_n =\frac{C_{-}A^{(0)}-\left(q^2T^{(0)}+q\varpi Z^{(0)}\right)}{1+i\varpi P(0)+q^2 Q(0)}
 \end{equation} 
and also, we have the expressions 
\begin{eqnarray}
&~&P(u)= -\frac{{\mathcal T}_0}{2T}\left(\frac{(1-u)\kappa^2}{(1+\kappa)(2+u\kappa)}+2 {\rm Log}\left[\frac{2+\kappa}{1+u+u\kappa}\right]\right)\nonumber\\
&~&Q(u)= \frac{\kappa^2(1-u)(2+\kappa)(2+3\kappa)}{2(1+\kappa)(2+u\kappa)(2+\kappa)^3}+\frac{2({\mathcal C}_1+{\mathcal D}_1)}{(2+\kappa)^3}\nonumber\\
&~&{\mathcal C}_1= \left(\frac{u\kappa^2(2+\kappa)}{2+u\kappa}+2(1+\kappa){\rm Log}[1+u+\kappa u]\right){\rm Log}[u]
+\kappa(2+\kappa){\rm Log}\left[\frac{2+\kappa}{1+u+u\kappa}\right]\nonumber\\
&~&{\mathcal D}_1=2(1+\kappa)\left({\rm Li}_2(1-u)- {\rm Li}_2(-1-\kappa) + {\rm Li}_2(-u(1+\kappa))\right)
\end{eqnarray}
where we have introduced ${\rm Li}$,  the PolyLogarithm function. Finally, by writing down the boundary action, we find in this case, upto leading order,
\begin{equation}
{\epsilon}=1-\frac{\pi C_{em}^2 N_{c}^{2}{\mathcal T}_0^2\kappa}{4\omega^{2}}+\frac{iC_{em}^2 N_{c}^{2}{\mathcal T}_0\left(2+\kappa\right)^2 }{16 \omega\sqrt{1+\kappa}}+....
\end{equation}
\begin{equation}
{\mu}=1+\frac{i C_{em}^2 N_{c}^2 {\mathcal T}_0\kappa}{16\omega \sqrt{1+\kappa}}+....
\end{equation}
\begin{figure}[t!]
\begin{minipage}[b]{0.5\linewidth}
\centering
\includegraphics[width=2.8in,height=2.3in]{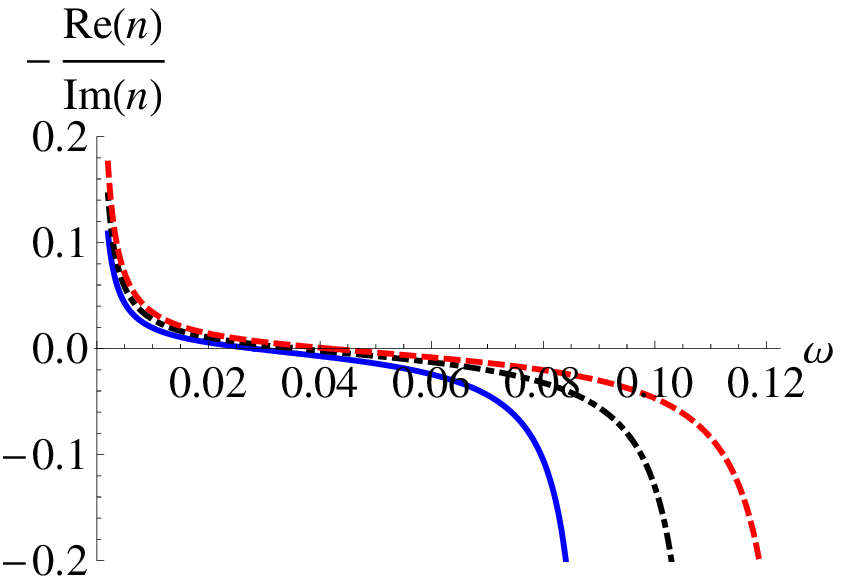}
\caption{$-\frac{Re[n]}{Im[n]}$ as a function of $\omega$, for
$\kappa=1$ (solid blue), 1.5 (dot dashed black) and 2 (dashed red).}
\label{fig11}
\end{minipage}
\hspace{0.2cm}
\begin{minipage}[b]{0.5\linewidth}
\centering
\includegraphics[width=2.8in,height=2.3in]{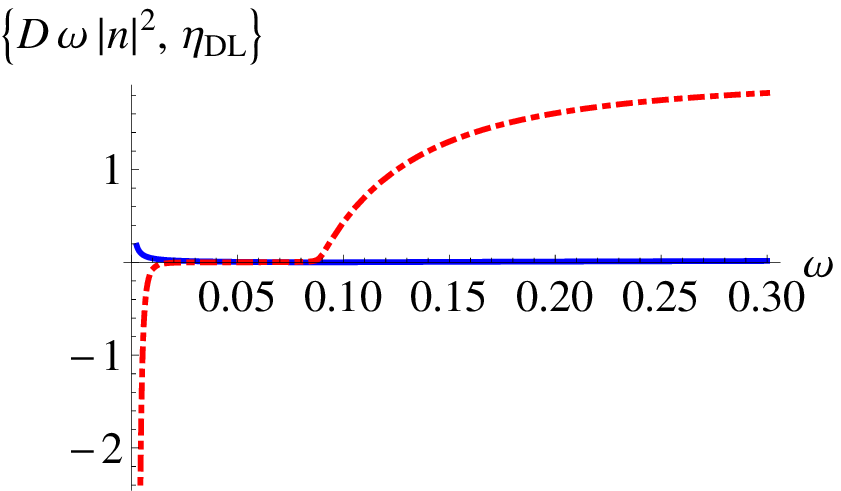}
\caption{$\omega D |n^2|$ (solid blue line) and $\eta_{DL}$ (dashed red line) as
functions of $\omega$, for $\kappa=1$.}
\label{fig12}
\end{minipage}
\end{figure}
Again we see that at small enough $\omega$, the real part of $\epsilon$ obtains negative values ensuring the occurrence of negative refractive index. 
The cut-off $\omega$ for negative refractive index will vary with the value of the charge parameter $\kappa$. To go beyond the hydrodynamic limit we have to
implement a numerical solution of eq.(\ref{main5d}). In figs. (\ref{5dImmu}) and (\ref{5dDL}) we show numerical results for the imaginary part of $\mu$ and the DL index, 
respectively. In figs. (\ref{fig11}) and (\ref{fig12}), we demonstrate the dissipation effects and the validity of the constraint $|n^2|\ll \frac{1}{\omega D}$
respectively. These are seen to have qualitatively similar features as the cases considered before.

\subsection{A Two-charge Example in 5-D}

We will now consider a two-charge example in five dimensions, with two non-zero chemical potentials. In this case, the metric and the gauge fields assume
the expressions 
\begin{eqnarray}
ds_5^{2} &=& \frac{\left(\pi  {\mathcal T}_0 L \right)^{2} }{u}H^{1/3}\left( \frac{-f}{H} dt^{2}+dx^{2}+dy^{2}+dz^{2}\right)+\frac{L^{2}}{4fu^{2}} H^{1/3}du^{2} \nonumber\\
A^{1}_{\mu } &=& \pi  {\mathcal T}_0\sqrt{2\kappa} \left( 1+\kappa \right)\frac{u}{H^{1/2}}\left(dt \right) _{\mu } \nonumber\\
A^{2}_{\mu } &=& \pi  {\mathcal T}_0\sqrt{2\kappa} \left( 1+\kappa \right)\frac{u}{H^{1/2}}\left(dt \right) _{\mu }
\end{eqnarray}
where we have defined 
\begin{equation}
H = H_{1}H_{2}=\left( 1 + \kappa u\right) ^2,~~~f = H-\left( 1+\kappa\right)^2 u^2
\end{equation}
It is enough for us to consider perturbations to the metric and gauge fields of the form 
\begin{eqnarray}
h_{tx} &=& g_{xx}\left( u\right) T\left( u\right) e^{-i\omega t + iKz},  h_{zx} =	g_{xx}\left( u\right) Z\left( u\right) e^{-i\omega t + iKz}\nonumber\\
a^{1}_{x} &=& \frac{\mu^1 }{2}A^1\left( u\right) e^{-i\omega t + iKz},a^{2}_{x} = \frac{\mu^2 }{2}A^2\left( u\right) e^{-i\omega t + iKz}
\end{eqnarray}
and we find that the linearized equations are given by 
\begin{eqnarray}
&~& T^{'} + \frac{qf}{\varpi H } Z^{'} + \frac{\kappa\left(1+\kappa\right)u}{2H}\left(A^1+A^2\right)  = 0
\nonumber\\
&~& T^{''} +\frac{uH^{'}- H}{u H }T^{'}-\frac{q^{2}}{fu }T-\frac{\varpi q}{ fu }Z +  \frac{\kappa\left(1+\kappa\right)u}{2H}\left({A^1}^{'}+{A^2}^{'}\right)  = 0
\nonumber\\
&~& Z^{''} + \frac{uf^{'}-f}{u f } Z^{'} + \frac{\varpi ^{2} H}{f^{2}u} Z + \frac{\varpi q H}{f^{2}u } T = 0
\nonumber\\
&~& {A^1}^{''} + \frac{f^{'}}{f } {A^1}{'} + \frac{2\left(1+\kappa\right)}{f} {T}{'}+\left(\frac{\varpi ^{2} H}{f^{2}u}-\frac{q^2}{uf}\right)A^1 = 0
\nonumber\\
&~& {A^2}^{''} + \frac{f^{'}}{f } {A^2}{'} + \frac{2\left(1+\kappa\right)}{f} {T}{'}+\left(\frac{\varpi ^{2} H}{f^{2}u}-\frac{q^2}{uf}\right)A^2 = 0
\end{eqnarray}
Where, $\varpi = \frac{\omega }{2\pi {\mathcal T}_0}$ and $q = \frac{K  }{2\pi {\mathcal T}_0}$. Now, we consider the new variables 
\begin{equation}
\Phi_{\pm }\left(u\right)=\frac{H}{u}T'\left(u\right) +\frac{\kappa(1+\kappa)}{2} \left( A^1\left(u\right)+A^2\left(u\right)\right) +C_{\pm}\frac{\sqrt{H}}{u}\left(A^1\left(u\right)+A^2\left(u\right)\right) 
\end{equation}
in which, we define 
\begin{equation}
C_{\pm}=\frac{1+\kappa}{4}\left(-1  \pm \sqrt{1+\frac{2q^2\kappa}{\left(1+\kappa\right)^2}}\right)
\end{equation}
and the perturbation equations decouple into the following form 
\begin{eqnarray}
&~&\Phi''_{\pm }\left(u\right) + \left(\frac{f'}{f}+\frac{2}{u}-\frac{H'}{H}\right) \Phi'_{\pm }\left(u\right)+\frac{1}{u f^2} \left(\varpi^2 H-q^2 f \right)
\Phi_{\pm }\left( u\right)+4C_{\pm} \frac{(1+\kappa)}{f \sqrt{H}} \Phi_{\pm }\left(u\right)=0
\nonumber\\
&~&\left(A^1-A^2\right)^{''} + \frac{f'}{f}\left(A^1-A^2\right)+\frac{1}{u f^2} \left(\varpi^2 H-q^2 f\right)\left(A^1-A^2\right)=0
\end{eqnarray}
Now, imposing the boundary conditions 
\begin{eqnarray}
&~&\lim _{u\rightarrow 0} \left[ u^2\Phi_{\pm }'\left(u\right)-C_{\pm}u\sqrt{H}\left( {A^1}^{'}+{A^2}^{'}\right)\right] =q^2T^{(0)}+q\varpi Z^{(0)}-C_{\pm}\left({A^1}^{0}+{A^2}^{0}\right) 
\nonumber\\
&~&{A^i}^{(0)}=\lim _{u\rightarrow 0} {A^i}\left(u\right),
T^{(0)}=\lim _{u\rightarrow 0} T\left(u\right),
Z^{(0)}=\lim _{u\rightarrow 0} Z\left(u\right)
\end{eqnarray} 
we can calculate the boundary action in the same way as done in the previous subsection. We omit the details of the calculation here. 
As in the previous single charge example, we now impose the incoming boundary wave conditions on $\Phi_+$, $\Phi_-$ and $(A^1-A^2)$ and solve for 
$A^1$, $A^2$, $T$ and $Z$ in the hydrodynamic limit using a series expansion in $\varpi $ and $q^2$. Then we put back these solutions in the boundary action and 
compute the retarded correlators. These are listed below: 
\begin{eqnarray}
&~& G_{xx}^{11}=\frac{i\omega N_{c}^{2}{\mathcal T}_0^{2}\kappa }{16\left(i\omega-\frac{K^2}{4\pi {\mathcal T}_0\left(1+\kappa\right)} \right)}-\frac{i\omega N_{c}^{2}{\mathcal T}_0
\left(2+3\kappa+2\kappa^2\right)}{32\pi\left(1+\kappa \right) }=G_{xx}^{22}
\nonumber\\
&~& G_{xx}^{12}=\frac{i\omega N_{c}^{2}{\mathcal T}_0^{2}\kappa }{16\left(i\omega-\frac{K^2}{4\pi {\mathcal T}_0\left(1+\kappa\right)} \right)}+\frac{i\omega N_{c}^{2}{\mathcal T}_0\left(3+2\kappa\right)\kappa}
{32\pi\left(1+\kappa \right) }=G_{xx}^{21}
\nonumber\\
&~& G_{xtxt}=\frac{\pi N_{c}^{2}{\mathcal T}_0^{3}\left(1+\kappa\right)K^2}{8\left(i\omega-\frac{K^2}{4\pi {\mathcal T}_0\left(1+\kappa\right)} \right)},~~~
G_{xzxz}=\frac{\pi N_{c}^{2}{\mathcal T}_0^{3}\left(1+\kappa\right)\omega^2}{8\left(i\omega-\frac{K^2}{4\pi {\mathcal T}_0\left(1+\kappa\right)} \right)}
\nonumber\\
&~& G_{xtxz}=-\frac{\pi N_{c}^{2}{\mathcal T}_0^{3}\left(1+\kappa\right)\omega K}{8\left(i\omega-\frac{K^2}{4\pi {\mathcal T}_0\left(1+\kappa\right)} \right)},~~~
G_{xtx}^{1}=\frac{i\pi N_{c}^{2}{\mathcal T}_0^{3}\left(1+\kappa\right)\sqrt{2\kappa}\omega}{8\left(i\omega-\frac{K^2}{4\pi {\mathcal T}_0\left(1+\kappa\right)} \right)}=G_{xtx}^{2}
\nonumber\\
&~& G_{xzx}^{1}=-\frac{ N_{c}^{2}{\mathcal T}_0^{2}\sqrt{2\kappa} \omega K}{32\left(i\omega-\frac{K^2}{4\pi {\mathcal T}_0\left(1+\kappa\right)} \right)}=G_{xzx}^{2}
\end{eqnarray}
The components of permittivity and permeability are computed from $G_{xx}^1$and $G_{xx}^2$ using the following relationship:
\begin{equation}
G_{xx}^{ij}=G_{xx0}^{ij}+K^2 G_{xx2}^{ij},~~{\epsilon}^{ij}=1-\frac{4\pi C_{em}^2}{\omega^{2}} G_{xx0}^{ij},~~{\mu}^{ij}=1-4\pi C_{em}^2 G_{xx2}^{ij}
\end{equation} 
And, they are found out to be,
\begin{eqnarray}
&~&{\epsilon}^{11}=1-\frac{\pi C_{em}^2 N_{c}^{2}{\mathcal T}_0^2\kappa}{4\omega^{2}}+\frac{iC_{em}^2 N_{c}^{2}{\mathcal T}_0\left(2+3\kappa+2\kappa^2\right) }
{8\left(1+\kappa\right)\omega}+....={\epsilon}^{22}
\nonumber\\
&~&{\epsilon}^{12}=1-\frac{\pi C_{em}^2 N_{c}^{2}{\mathcal T}_0^2\kappa}{4\omega^{2}}-\frac{iC_{em}^2 N_{c}^{2}{\mathcal T}_0\kappa\left(3+2\kappa\right) }
{8\left(1+\kappa\right)\omega}+....={\epsilon}^{21}
\nonumber\\
&~&{\mu}^{11}=1+\frac{i C_{em}^2 N_{c}^2 {\mathcal T}_0\kappa}{16\left(1+\kappa\right)\omega }+....={\mu}^{22}={\mu}^{12}={\mu}^{21}
\end{eqnarray} 
As before, we have numerically analyzed the DL index using the above relations, away from the hydrodynamic limit. We find that this has the
same qualitative features as the examples considered till now, namely that the DL index becomes negative at sufficiently small values of the 
frequency. Having the master variables in this case again allows for the computation of all the correlators away from the hydrodynamic limit, as
indicated previously. 

\section{7d Single R-charged black hole}

Finally, we turn to the case of R-charged black holes in seven dimensions, which correspond to rotating $M5$-branes. The analysis of the momentum dependent
vector modes is similar to what we have described, and we will be brief here. The metric and the gauge fields are given by 
\begin{eqnarray}
ds_7^{2} &=& \frac{4\left(\pi  {\mathcal T}_0 L \right)^{2} }{9u}H^{1/5}\left( \frac{-f}{H} dt^{2}+dx_{1}^{2}+.....+dx_{4}^{2}+dz^{2}\right)+\frac{L^{2}}{4fu^{2}} H^{1/5}du^{2} 
\nonumber\\
A_{\mu } &=& \frac{2}{3}\pi  {\mathcal T}_0\sqrt{2\kappa \left( 1+\kappa \right) }\frac{u^2}{H}\left(dt \right) _{\mu }
\end{eqnarray}
Where we define
\begin{equation}
H = 1 + \kappa u^2, ~~~f = \left(1-u \right)\left\lbrace 1 +u +  \left(1+\kappa  \right)u^{2} \right\rbrace,~~~
\frac{T}{{\mathcal T}_0} = \frac{1 + \frac{\kappa }{3}}{\sqrt{1+\kappa }}
\end{equation}
where, as before, $\kappa$  denotes the R-charge while ${\mathcal T}_0$ is the Hawking temperature of the neutral black hole \cite{natsuume}.
The perturbations are as in the previous examples, and we obtain the linearized Einstein's and Maxwell's equation in terms of the perturbations as : 
\begin{eqnarray}
&~& A''\left(u\right) +\left( \frac{f'}{f}+\frac{H'}{H}-\frac{1}{u}\right)A'\left( u\right) +\frac{4\left(1+\kappa \right)u }{f H}T'\left( u\right) +\frac{9}{4}\left(\frac{\varpi ^2 H}{uf^2}
-\frac{q^2}{uf} \right)A\left( u\right) =0  
\nonumber\\
&~& Z''\left(u\right) + \frac{uf'-2f}{uf}Z'\left(u\right) +\frac{9}{4}\frac{\varpi ^2 H}{uf^2} Z\left(u\right) +\frac{9}{4}\frac{\varpi q H}{uf^2}T\left( u\right) =0  
\nonumber\\
&~& T''\left(u\right) + \frac{uH'-2H}{uH}T'\left( u\right) - \frac{9}{4}\frac{q^2}{uf}T\left(u\right) - \frac{9}{4}\frac{\varpi q}{uf} Z\left(u\right) + \frac{\kappa u^2}{H}A'\left(u\right)   =0  
\nonumber\\
&~& T'\left(u\right) + \frac{q f}{\varpi H}Z'\left( u\right) + \frac{\kappa u^2}{H}A\left(u\right)   =0  
\end{eqnarray}
Where,
\begin{equation}
H=1+\kappa u^2,~~f=H-(1+\kappa)u^3,~~\varpi=\frac{\omega}{2\pi {\mathcal T}_0},~~q=\frac{K}{2\pi {\mathcal T}_0}
\end{equation}
The master variables are constructed as 
\begin{equation}
\Phi_{\pm }\left(u\right)=\frac{H}{u^2}T'\left(u\right) +\kappa A\left(u\right) +C_{\pm}\frac{H}{u^2}A\left(u\right)
\end{equation}
Where we need to use the definition
\begin{equation}
C_{\pm}=\frac{3}{4 }\left(-1+\sqrt{1+\frac{q^2\kappa}{2\left(1+\kappa\right)}}\right)
\end{equation}
In terms of the master variable, we get the e.o.m as 
\begin{eqnarray}
\Phi''_{\pm }\left(u\right) &+& \left(\frac{f'}{f}+\frac{3}{u}-\frac{H'}{H}\right) \Phi'_{\pm }\left(u\right)+\frac{9}{4uf^2} \left(\varpi^2 H-q^2 fu \right)\Phi_{\pm }\left( u\right)
\nonumber\\
&+& C_{\pm} \frac{4(1+\kappa)u}{f H} \Phi_{\pm }\left(u\right)=0
\end{eqnarray}
We will simply present the results of the permittivity and the effective permeability, which in the hydrodynamic limit can be shown to be 
\begin{eqnarray} 
\epsilon &=& 1-\frac{ 256 \pi^2 C_{em}^2 N_{c}^3 {\mathcal T}_0^{4} \kappa }{729\omega ^2}+\frac{16i C_{em}^2 N_{c}^{3}\pi {\mathcal T}_0^{3} (\kappa +3)^2 }{729 \omega \sqrt{\kappa +1}}+.....
\nonumber\\
\mu &=& 1+\frac{64\pi i C_{em}^2 N_{c}^{3} {\mathcal T}_0^3\kappa}{729\sqrt{3+\kappa}\omega}+.....
\end{eqnarray} 
A numerical method is now used to obtain these expressions beyond the hydrodynamic limit, and as before, we find that the behavior of the DL index is
similar to the four and five dimensional cases, and that this index becomes negative for sufficiently small values of the frequency. 

\section{Discussions and Conclusions}

In this paper, we have studied momentum dependent vector mode perturbations for R-charged black holes. We have carried out this analysis in
four, five and seven dimensions, for both single and two-charged black holes. We have explicitly constructed the master variables in each case, for
which the perturbation equations decouple. In principle, this allows for numerical analysis of the correlators away from the hydrodynamic limit. 
In particular, in order to study optical properties of boundary theories dual to these black holes, we have used our results to evaluate the DL index. 
In all the cases considered here, we find that the index becomes negative at sufficiently small frequencies, confirming
the prediction of \cite{policastro1}. 

Here, we have only considered the case of the flat horizon. It will be very interesting to extend our analysis to the case of spherical horizons in 
R-charged black holes. This might shed light on the behavior of correlation functions away from the hydrodynamic limit, near critical points. 
Further, we did not construct the master variables for generic multi charge black holes. Our two-charge examples here are limited to the
case where the charges are equal. Work is in progress in this direction. 

\begin{center}
{\bf Acknowledgements}
\end{center}
It is a pleasure to thank S-J. Sin and Sachin Jain for useful email correspondence.


\begin{thebibliography}{999}
\bibitem{malda}
O.~Aharony, S.~S.~Gubser, J.~M.~Maldacena, H.~Ooguri and Y.~Oz,
``Large N field theories, string theory and gravity,''
Phys.\ Rept.\  {\bf 323}, 183 (2000) {\tt [hep-th/9905111]}.
\bibitem{hartnoll}
S.~A.~Hartnoll,
``Lectures on holographic methods for condensed matter physics,'' Class.\ Quant.\ Grav.\  {\bf 26}, 224002 (2009) {\tt [arXiv:0903.3246 [hep-th]]}.
\bibitem{sachdev1}
S. Sachdev, 
``Condensed Matter and AdS/CFT,''
Lect.\ Notes Phys.\  {\bf 828}, 273 (2011), {\tt [arXiv:1002.2947 [hep-th]]}.
\bibitem{matsuo1}
Y.~Matsuo, S.~-J.~Sin, S.~Takeuchi, T.~Tsukioka and C.~-M.~Yoo,
``Sound Modes in Holographic Hydrodynamics for Charged AdS Black Hole,'' Nucl.\ Phys.\ B {\bf 820}, 593 (2009), {\tt [arXiv:0901.0610 [hep-th]]}.
\bibitem{matsuo2}
Y.~Matsuo, S.~-J.~Sin, S.~Takeuchi and T.~Tsukioka,
``Magnetic conductivity and Chern-Simons Term in Holographic Hydrodynamics of Charged AdS Black Hole,'', JHEP {\bf 1004}, 071 (2010),
{\tt [arXiv:0910.3722 [hep-th]]}.
\bibitem{leigh}
M.~Edalati, J.~I.~Jottar and R.~G.~Leigh, 
``Shear Modes, Criticality and Extremal Black Holes,'' JHEP {\bf 1004}, 075 (2010), {\tt [arXiv:1001.0779 [hep-th]]}.
\bibitem{kodamaishibashi}
H.~Kodama and A.~Ishibashi,
``Master equations for perturbations of generalized static black holes with charge in higher dimensions,''
Prog.\ Theor.\ Phys.\  {\bf 111}, 29 (2004), {\tt [arXiv:hep-th/0308128]}.
\bibitem{sar}
S. A. Ramakrishna, ``Physics of negative refractive index materials,'' Rep. Prog. Phys. {\bf 68} (2005) 449. 
\bibitem{valesagorev}
V. Vaselago, L. Braginsky, V. Shklover, C. Hafner, Journ. Comput. Theor. Nanoscience, {\bf 3} (2006) 189.
\bibitem{veselago}
V. G. Veselago, ``The electrodynamics of substances with simultaneously negative values of $\epsilon$ and $\mu$, 
Sov. Phys. Usp. {\bf 10} (1968) 509. 
\bibitem{pendry1}
J. B. Pendry, A. J. Holden, W. J. Stewart, I. Youngs, ``Extremely low frequency plasmons in metallic mesostructures,'' Phys. Rev. Lett. 
{\bf 76} (1996) 4773.
\bibitem{pendry2}
J. B. Pendry, A. J. Holden, D. J. Robbins, W. J. Stewart, `` Magnetism from conductors and enhanced nonlinear phenomena,'' 
IEEE Trans. Microwave Theory and Techniques, {\bf 47} (1999) 2075. 
\bibitem{exp}
D. R. Smith, W. J. Padilla, D. C. Vier, S. C. Nemat-Nasser, S. Schultz, ``Composite medium with simultaneously negative permeability and permittivity,''
Phys. Rev. Lett. {\bf 84} (2000) 4184.
\bibitem{policastro1}
A. Amariti, D. Forcella, A. Mariotti, G. Policastro, ``Holographic optics and negative refractive index,'' 
JHEP {\bf 1104}, (2011) 036, {\tt [arXiv:1006.5714 [hep-th]]}.
\bibitem{sin1}
X.~-H.~Ge, K.~Jo and S.~-J.~Sin,
 ``Hydrodynamics of RN AdS$_4$ black hole and Holographic Optics,''
 JHEP {\bf 1103}, 104 (2011), {\tt [arXiv:1012.2515 [hep-th]]}.
\bibitem{hartnoll2}
S.~A.~Hartnoll, C.~P.~Herzog and G.~T.~Horowitz,
``Building a Holographic Superconductor,''
Phys.\ Rev.\ Lett.\  {\bf 101}, 031601 (2008), {\tt [arXiv:0803.3295 [hep-th]]}.
\bibitem{gaozhang}
X.~Gao and H.~-b.~Zhang,
``Refractive index in holographic superconductors,''
JHEP {\bf 1008}, 075 (2010), {\tt [arXiv:1008.0720 [hep-th]]}.
\bibitem{policastro2}
 A.~Amariti, D.~Forcella, A.~Mariotti and M.~Siani,
 ``Negative Refraction and Superconductivity,''
JHEP {\bf 1110}, 104 (2011), {\tt[arXiv:1107.1242 [hep-th]]}.
\bibitem{natsuume}
K.~Maeda, M.~Natsuume and T.~Okamura,
 ``Dynamic critical phenomena in the AdS/CFT duality,'' Phys.\ Rev.\ D {\bf 78}, 106007 (2008) {\tt [arXiv:0809.4074 [hep-th]]}.
\bibitem{gubser}
 S.~S.~Gubser,
 ``Thermodynamics of spinning D3-branes,''
Nucl.\ Phys.\ B {\bf 551}, 667 (1999), {\tt [arXiv:hep-th/9810225]}.
\bibitem{cvetic}
K.~Behrndt, M.~Cvetic and W.~A.~Sabra,
 ``Nonextreme black holes of five-dimensional N=2 AdS supergravity,'' Nucl.\ Phys.\ B {\bf 553}, 317 (1999) {\tt [arXiv:hep-th/9810227]},\\
 M.~Cvetic et al., ``Embedding AdS black holes in ten-dimensions and eleven-dimensions,'' Nucl.\ Phys.\ B {\bf 558}, 96 (1999), {\tt [arXiv:hep-th/9903214]}.
\bibitem{sonstarinets}
D.~T.~Son and A.~O.~Starinets,
``Hydrodynamics of r-charged black holes,'' JHEP {\bf 0603}, 052 (2006), {\tt [arXiv:hep-th/0601157]}.
\bibitem{sachin1}
S.~Jain, S.~Mukherji and S.~Mukhopadhyay,
``Notes on R-charged black holes near criticality and gauge theory,'' JHEP {\bf 0911}, 051 (2009), {\tt [arXiv:0906.5134 [hep-th]]}.
\bibitem{sachin2}
 S.~Jain, 
 ``Universal properties of thermal and electrical conductivity of gauge theory plasmas from holography,'' JHEP {\bf 1006}, 023 (2010), {\tt [arXiv:0912.2719 [hep-th]]}.
\bibitem{deplak}
R. A. Depine, A. Lakhtakia, ``A new condition to identify isotropic dielectric-magnetic materials displaying negative phase velocity,''
Micro. Opt. Tech. Lett. {\bf 41} (2004) 315. 
\end{thebibliography}
\end{document}